\documentclass[letter]{article}
\usepackage[total={17.75cm,24.5cm},top=1.5cm,left=2.0cm]{geometry}
\usepackage{xspace,verbatim,graphicx}
\usepackage{amsmath,amssymb}
\usepackage{color,cite,authblk}
\usepackage[squaren]{SIunits}
\usepackage[hidelinks]{hyperref}
\usepackage{lineno}

\newcommand{\me}[1]{\ensuremath{\mathrm{e}^{#1}}\xspace}
\newcommand{\ddt}[1]{\frac{\mathrm{d}#1}{\mathrm{d}t}}

\newcommand{\Cbar}{\ensuremath{\kappa}\xspace}
\newcommand{\Npts}{\ensuremath{N_\text{pts}}\xspace}

\newcommand{\pvec}{\ensuremath{\vec{p}}\xspace}

\newcommand{\posterior}{\ensuremath{\text{Posterior}}\xspace}
\newcommand{\likeli}{\ensuremath{\mathcal{L}}\xspace}
\newcommand{\prior}{\ensuremath{\text{Prior}}\xspace}

\title{A closer look at parameter identifiability, model selection and handling of censored data with Bayesian Inference in mathematical models of tumour growth}

\author[1]{Jamie Porthiyas}
\author[1]{Daniel Nussey}
\author[2,3]{Catherine A.\ A.\ Beauchemin}
\author[3]{Donald C.\ Warren}
\author[2]{Christian Quirouette}
\author[1]{Kathleen P.\ Wilkie}

\affil[1]{Department of Mathematics, Toronto Metropolitan University, Toronto, ON, M5B 2K3, Canada}
\affil[2]{Department of Physics, Toronto Metropolitan University, Toronto, ON, M5B 2K3, Canada}
\affil[3]{Interdisciplinary Theoretical and Mathematical Sciences (iTHEMS) program, RIKEN, Wako-shi, Saitama, 351-0198, Japan}

\date{\today}

\begin{document}

\maketitle

\begin{abstract}
Mathematical models (MMs) are a powerful tool to help us understand and predict the dynamics of tumour growth under various conditions. In this work, we use 5 MMs with an increasing number of parameters to explore how certain (often overlooked) decisions in estimating parameters from data of experimental tumour growth affect the outcome of the analysis. In particular, we propose a framework for including tumour volume measurements that fall outside the upper and lower limits of detection, which are normally discarded. We demonstrate how excluding censored data results in an overestimation of the initial tumour volume and the MM-predicted tumour volumes prior to the first measurements, and an underestimation of the carrying capacity and the MM-predicted tumour volumes beyond the latest measurable time points. We show in which way the choice of prior for the MM parameters can impact the posterior distributions, and illustrate that reporting the highest-likelihood parameters and their 95\% credible interval can lead to confusing or misleading interpretations. We hope this work will encourage others to carefully consider choices made in parameter estimation and to adopt the approaches we put forward herein.
\end{abstract}

%%%%%%%%%%%%%%%%%%%%%%%%%%%%%%%%%%%%%%%%%%%%%%%%%%%%%%%%%%%%%%%%%%%%%%%%%%%%%%%%
%%%%%%%%%%%%%%%%%%%%%%%%%%%%%%%%%%%%%%%%%%%%%%%%%%%%%%%%%%%%%%%%%%%%%%%%%%%%%%%%
%%%%%%%%%%%%%%%%%%%%%%%%%%%%%%%%%%%%%%%%%%%%%%%%%%%%%%%%%%%%%%%%%%%%%%%%%%%%%%%%
\cleardoublepage
\section{Introduction}

Mathematical models (MM) are the primary tools by which we can examine biological or clinical data to determine fundamental mechanisms, to test hypotheses, and to make predictions. A MM formalizes assumptions of causality with the aim of exploring the limits and consequences of the input-output relationship. In the process of formalizing biological processes and function into mathematical forms, parameter values are introduced. The values of these parameters then become key factors in the MM's predictions, and are typically estimated by fitting the MM to data (least-square approach) or estimating their likelihood given the observed data (Bayesian approach). Determining MM parameters given an observed outcome is a challenging inverse problem that is highly sensitive to noise. 

MMs can be used to predict an average response (by using a set of parameter values based on all trajectories at once or on their average), to predict a specific response (by using a set of parameter values based on a specific trajectory in a dataset), and to predict population-level responses (by using an ensemble of parameter value sets obtained in a manner that generates parameter joint-distributions). Prediction of a patient-specific response requires estimating based on a specific trajectory, and can be used to develop a digital-twin if the estimation is repeated to update parameter values as new data is obtained \cite{Stahlberg2022}.  Population responses require the MM to be extended by a virtual clinical trial where response is tracked for a virtual population represented by an ensemble of parameter value sets \cite{Craig2023}.  

A MM is classified as identifiable if the parameters can be uniquely determined given the system inputs and outputs \cite{Phan2023}. In practice, this requires comparing the MM outputs to noisy real-world data \cite{Eisenberg2017}. Simply put, if two different sets of MM parameters exist that give identical MM dynamics, then the MM is not identifiable. A MM may not be practically identifiable for two main reasons: 1) the data is not sufficient to identify all MM parameters (for example, estimating the parameters of a logistic growth curve based on data that only demonstrates an exponential growth phase), and 2) two or more MM parameters are somehow dependent or their actions are coupled as measured by the MM output (for example, estimating the parameters of an exponential growth MM with a net growth rate of $a-b$ where $a$ is the growth rate and $b$ is the death rate --- only the difference $a-b$ is identifiable). For MM predictions to be well constrained beyond the data, the MM taken together with the dataset should be practically identifiable. Ways to improve identifiability generally include either reducing the number of MM parameters and/or collecting and using additional data such as more data points over an extended time range, or additional input conditions such as various doses of the given treatment.

Despite best efforts, however, it is often the case that a MM coupled with all available data, which is often extremely limited in clinical applications, is not practically identifiable.  In this case, the parameter estimation process may result in many equivalently highly likely (or best-fit) parameter sets. Off hand, this may be taken as a negative, since non-identifiability can cause the MM to produce vastly different outputs to the same input conditions for these various equally likely parameter sets. It can, however, also be a considerable positive, if the aim of the modelling is to capture variability and heterogeneity in response \cite{Wilkie2013, Wilkie2017}, such as in a virtual clinical trial. The multiple accepted parameter sets form an ensemble that can represent a virtual cohort, and the differing response dynamics then extend the capabilities of the MM to explore the resulting heterogeneity \cite{Craig2023}.

For example, biological factors such as the sensitivity to angiogenic signals and the timescale associated with the sprouting and development of new blood vessels can affect the dynamic carrying capacity of a tumour microenvironment. This carrying capacity, modelled as either a constant parameter or a dependent variable, plays a significant role in determining the dynamic behaviour of a growing tumour and its responsiveness to treatment \cite{hahnfeldt99, Wilkie2013, Wilkie2017}. Unfortunately, such factors are inherently difficult to estimate from data as no measurement can be taken directly on the capacity of a microenvironment to support a tumour. In a logistic-{} or Gompertz-type MM, the ratio of tumour volume to carrying capacity slows down exponential growth until the maximum size is obtained. In the generalized logistic MM, there is an additional parameter that controls the strength of this ratio's effect on tumour growth. The tumour volume to carrying capacity ratio is so significant, it was coined the proliferation saturation index in radiation response modelling \cite{Prokopiou2015}, and was shown to play a significant role in determining patient-specific responses to radiation in a MM where the radiation effect directly altered the carrying capacity \cite{Zahid2021}.

The best situation to parameterize a MM is to have complete time-series datasets for every output in the MM as well as additional insights or data on mechanisms described by the MM. Then, the MM can be parameterized and validated fully before being used to explore alternative situations from those described by the data~\cite{Brady2019}. If the aim is to study competing hypotheses for cancer development and treatment using MMs, then it is important that the parameters and/or the MM-predicted tumour growth curve beyond the extent of the data be sufficiently constrained by their estimation from available data, in order to challenge and discriminate between competing hypotheses. If a MM and dataset taken together are identifiable, then the region of highest likelihood in the parameter space will be well constrained and the resulting MM predictions will be as well.

On the other hand, increasingly complex MMs that capture the biological processes in greater detail, perhaps in order to correctly capture the mechanism behind a particular therapy, will lead to more parameters. This increase will likely ensure that the MM and data together are not identifiable, making the higher-dimensional parameter landscape more complex, with potentially many local minima, or disconnected parameter space regions of equivalently high likelihood. In such cases, a richer data set is required to adequately constrain the values of the additional MM parameters, although such data might not be available or even obtainable. Within the poorly constrained high likelihood regions of the parameter space, MM predictions could differ significantly and affect conclusions, if not realized and handled with care.

Here, we consider several MMs of tumour growth with increasing complexity, and thus, number of parameters. We use Bayesian inference to examine the ability of an experimental data to constrain each MM's parameters, and explore the ability of the resulting parameterizations to predict growth beyond the measured time points. Importantly, we highlight the effects of missing data points and the choice of prior on the MM parameterization results.

%\cleardoublepage
%%%%%%%%%%%%%%%%%%%%%%%%%%%%%%%%%%%%%%%%%%%%%%%%%%%%%%%%%%%%%%%%%%%%%%%%%%%%%%%%
%%%%%%%%%%%%%%%%%%%%%%%%%%%%%%%%%%%%%%%%%%%%%%%%%%%%%%%%%%%%%%%%%%%%%%%%%%%%%%%%
%%%%%%%%%%%%%%%%%%%%%%%%%%%%%%%%%%%%%%%%%%%%%%%%%%%%%%%%%%%%%%%%%%%%%%%%%%%%%%%%
\section{Methods}

%%%%%%%%%%%%%%%%%%%%%%%%%%%%%%%%%%%%%%%%%%%%%%%%%%%%%%%%%%%%%%%%%%%%%%%%%%%%%%%%
\subsection{Considering a range of tumour growth MMs}

Let us first consider a modified, special case of the generalised logistic growth equation, also known as Richards' curve
\begin{align}
\ddt{C} &= \frac{\mu}{\min(\alpha,1)}\ C\ \left[1 - \left(\frac{C}{\Cbar}\right)^{\alpha} \right] \nonumber &
C(t) &= \frac{\Cbar}{\left[\ 1 + \left\{\left(\Cbar/C_0\right)^{\alpha} - 1\right\} \me{-\max(1,\alpha) \mu t}\ \right]^{(1/\alpha)}}
\tag{Rich MM} \label{MMgen}
\end{align}
where $C(t)$ is the tumour volume in $\milli\meter^3$ (wherein $\unit{1}{\milli\meter^3} \approx \unit{10^6}{Lewis}$ lung carcinoma (LLC) cells), and $\Cbar$ is both the fixed carrying capacity and the steady state of $C(t\to+\infty)$. The coefficient $\mu/\min(\alpha,1)$ is either $\mu$ for $\alpha\ge1$ or $\mu/\alpha$ for $\alpha<1$. This seemingly peculiar choice of coefficient better handles the change in the behaviour of this function about $\alpha=1$.

The \ref{MMgen} simplifies to the Logistic growth equation for $\alpha=1$,
\begin{align}
\ddt{C} &= \mu\ C\ \left[1 - \frac{C}{\Cbar} \right] \nonumber &
C(t) &= \frac{\Cbar}{1 + \left[\Cbar/C_0 - 1\right] \me{-\mu t}} \ ,
\tag{Logis MM} \label{MMlogis}
\end{align}
to the Gompertz growth equation as $\alpha\to0$,
\begin{align}
\ddt{C} &= - \mu\ C\ \ln\left[ \frac{C}{\Cbar} \right] \nonumber &
C(t) &= \Cbar \left[ \frac{C_0}{\Cbar} \right]^{\me{-\mu t}} \ ,
\tag{Gomp MM} \label{MMgomp}
\end{align}
and to exponential growth capped at $\Cbar$ as $\alpha\to+\infty$, namely
\begin{align}
C(t) &\approx \min(C_0 \me{\mu t}, \Cbar) \ .
\tag{ExpCap MM} \label{MMexpcap}
\end{align}
As the simplest MM we consider unbounded, exponential growth of the tumour, expressed as
\begin{align}
\ddt{C} &= \mu\ C &
C(t) = C_0 \me{\mu t} \ .
\tag{Exp MM} \label{MMexp}
\end{align}
The \ref{MMgen} has 4 unknown quantities ($\Cbar$, $\mu$, $\alpha$, $C_0$) to be estimated, the \ref{MMlogis}, \ref{MMgomp} and \ref{MMexpcap} have 3 ($\Cbar$, $\mu$, $C_0$), and the \ref{MMexp} has 2 ($\mu$, $C_0$).

%%%%%%%%%%%%%%%%%%%%%%%%%%%%%%%%%%%%%%%%%%%%%%%%%%%%%%%%%%%%%%%%%%%%%%%%%%%%%%%%
\subsection{Estimation of parameter posterior likelihood distributions}

For each MM variant considered, parameters' likelihood function \likeli, given by Eqns.\ \eqref{MMlike} or \eqref{MMlikeLOD}, and their associated \posterior was estimated using the MCMC method implemented by \texttt{phymcmc} \cite{phymcmc}, a graphing and analysis wrapper for \texttt{emcee} \cite{emcee}. The estimated \posterior{s} correspond to 300 chains of 10,000 steps each, yielding 3,000,000 parameter sets, which is preceded by a burn-in of no less than 10,000 steps. The process yields \unit{3}{M} accepted parameter sets, of which at least \unit{30}{k} (1\%) are completely independent based on the computed autocorrelation time \cite{emcee}, which gives a rough approximation of how many MCMC steps must separate a chain's past and present positions (parameter values) in order for the two positions to no longer be correlated.

%%%%%%%%%%%%%%%%%%%%%%%%%%%%%%%%%%%%%%%%%%%%%%%%%%%%%%%%%%%%%%%%%%%%%%%%%%%%%%%%
%%%%%%%%%%%%%%%%%%%%%%%%%%%%%%%%%%%%%%%%%%%%%%%%%%%%%%%%%%%%%%%%%%%%%%%%%%%%%%%%
%%%%%%%%%%%%%%%%%%%%%%%%%%%%%%%%%%%%%%%%%%%%%%%%%%%%%%%%%%%%%%%%%%%%%%%%%%%%%%%%
\cleardoublepage
\section{Results}

%%%%%%%%%%%%%%%%%%%%%%%%%%%%%%%%%%%%%%%%%%%%%%%%%%%%%%%%%%%%%%%%%%%%%%%%%%%%%%%%
\subsection{Important considerations in parameter estimation}

The experimental data considered herein corresponds to the control group in data published by and described in Benzekry et al.\ \cite{benzekry17}. Briefly, ten C57BL6 mice were injected sub-cutaneously, on the caudal-half of their back, with $10^6$ Lewis Lung Carcinoma (LLC) cells, said to correspond to a tumour volume of $\sim\unit{1}{\milli\metre^3}$. Measurements were taken by calipers and recorded in $\milli\metre^3$ at various times post injection. Due to the small tumour volumes shortly after injection, only 2/10 mice could be measured at 5 days post-injection (dpi), 8/10 at 6 dpi and 7 dpi, and 10/10 from then on. At later times, mice were euthanized for ethical reasons once tumours reached a maximum volume of \unit{1.5}{\centi\metre^3} such that 9/10 remained at 18 dpi, 7/10 at 19 dpi, 5/10 at 20 dpi, 2/10 at 21 dpi, and only 1/10 remained at 22 dpi.

We used a Markov chain Monte Carlo (MCMC) method to sample and ultimately estimate the posterior likelihood distribution (hereafter \posterior) of each MM's parameters from Bayes theorem. The \posterior of the MM parameter set \pvec, given the experimental data, is given by
\begin{align}
\posterior(\pvec|\mathrm{data}) =
\frac{\mathcal{L}(\mathrm{data}|\pvec) \cdot \prior(\pvec)}{\mathcal{P}(\text{data})} \propto \mathcal{L}(\mathrm{data}|\pvec) \cdot \prior(\pvec)\ .
  \label{MMpost}
\end{align}
For the likelihood of the data given \pvec, we first consider simply
\begin{align}
\mathcal{L}(\mathrm{data}|\pvec)
= \exp\left[-\frac{\mathrm{SSR}(\pvec)}{2\,\sigma_C^2}\right]
= \exp\left[-\frac{\sum_{k=1}^{\Npts} \sum_{\text{mouse}=1}^{\le10} \left\{\log_{10}[C_\text{MM}(\pvec,t_k)]- \log_{10}[C_{\text{mouse}}(t_k)]\right\}^2}{2\,\sigma_C^2}\right]\ ,
\label{MMlike}
\end{align}
where SSR is the sum of squared residuals between the $\log_{10}$ of the MM-predicted tumour volume at time point $t_k$ given \pvec, $C_\mathrm{MM}(\pvec,t_k)$, and that observed in each mouse for which that time point was measurable, $C_{\text{mouse}}(t_k)$.
The variance of the data, $\sigma_C^2$, is fixed, as described below.

\begin{figure}[htbp!]
\centering
\includegraphics[width=0.65\textwidth]{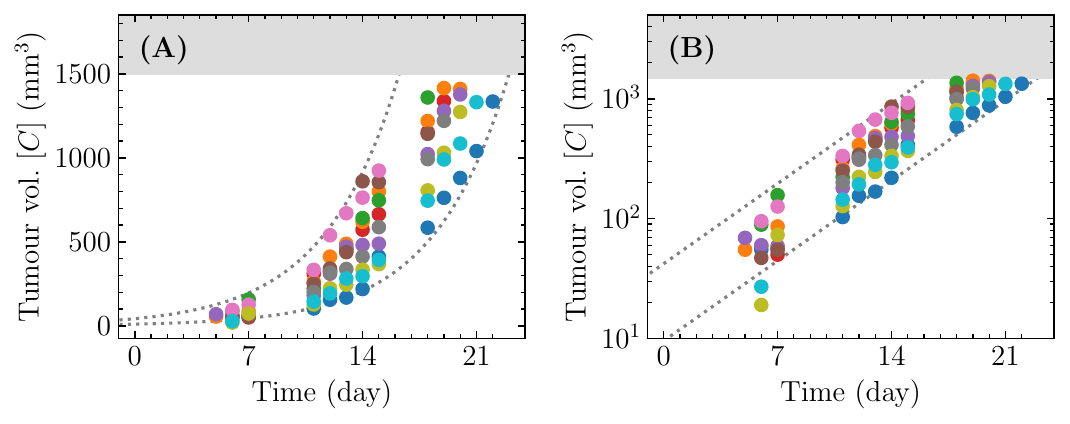}
\caption{%
\textbf{Experimentally measured tumour volume over time.}
The different coloured dots represent the tumour volume time course of 10 distinct mice from \cite{benzekry17}, shown using either a (A) linear or (B) logarithmic $y$-axis. The upper grey band corresponds to the upper limit of detection ($C=\unit{1.5\times10^3}{mm^3}$). The dashed grey lines correspond to the \ref{MMexp}-predicted tumour volume ($C_0=\unit{20}{mm^3}$, $\mu=\unit{0.22}{d^{-1}}$) multiplied or divided by 2.1 ($10^{2\sigma_C}$), matching the increasing spread in measurements for larger volumes. This indicates inter-mouse variability in $C$ is consistent with a log-normal distribution, following Eqn.\ \eqref{MMlike}.
}
\label{fig:bestfit}
\end{figure}

The likelihood of a parameter set was chosen to be a function of the residuals of $\log_{10}(C)$, rather than the tumour volume, $C$. This is because the inter-mouse variability in the tumour volume data is more consistent with a log-normal ($C^{\times}_{\div}\text{err}$) rather than a normal ($C\pm\text{err}$) distribution. This can be seen in Figure \ref{fig:bestfit} where the data is shown using both linear and logarithmic scales. The $\log_{10}$ tumour volume measurements are symmetrically distributed about the mean $\log_{10}(C)$ at each time point with a constant standard deviation, i.e.\ one that does not depend on $\log_{10}C$. Alternatively, one could compute the SSR for $C$ rather than $\log_{10}(C)$ by placing weights on tumour volume residuals that are proportional to $1/C$.

Correctly choosing the weights of residuals between model and data is an important step to perform before proceeding with parameter estimation. Yet, this step is often overlooked and data variability is assumed to be normally rather than log-normally distributed. In our experience, the latter error distribution is far more commonly encountered in experimental measurements in biology. Indeed, for the specific case of tumour growth, Benzekry et al.\ \cite{benzekry14,vaghi20} have demonstrated that the assumption of normal error distribution is incorrect for measurement of growing tumour volume over time.

A fixed standard deviation $\sigma_C=0.16$ was used for all data points, which corresponds to the standard deviation of $\log_{10}[C_\text{mouse}(t_k)]$ across all mice at time point $t_k$, averaged over all $t_k$. This means that at any given time point $t_k$, 95\% of the $\log_{10}$ tumour volume measurements should fall within $\pm2\sigma_C$ of the mean $\log_{10}(C)$ at that time point, or about 2.1-fold ($10^{2\sigma_C}$), shown as dashed grey lines in Figure \ref{fig:bestfit}.

Parameter estimation was performed using tumour volume measurements for all mice with measurable tumour volumes over all time points, rather than based on the average tumour volume at each time point. This implicitly takes into consideration the variable number of measurements at each time point, avoiding the issues of having to weigh certain measurement time points more or less heavily based on the number of points they represent, or to decide how best to average measurements (e.g., arithmetic vs geometric average) at each time point. Importantly, the process of handling all points as a single set is mathematically equivalent to estimating the parameters' \posterior for the first mouse, subsequently using it as the prior in determining the \posterior for the second mouse, and so on. This is, in essence, what some would call Machine Learning.

The parameter sets explored by the MCMC runs were used not only to sample and estimate the MM parameters' \posterior, but also to efficiently sample the shape of the likelihood function around the best-fit and/or the most likely parameter set. Herein, the \emph{best-fit} parameter set refers to that which maximizes the likelihood function, Eqn.\ \eqref{MMlike}, whereas the \emph{most likely} parameter set is that which maximizes the posterior distribution, Eqn.\ \eqref{MMpost}. This distinction, and specifically the Prior and Posterior distributions in Eqn.\ \eqref{MMpost}, will be discussed in more detail in later sections. Initially, we will focus on the likelihood function and the best-fit parameters alone.

%%%%%%%%%%%%%%%%%%%%%%%%%%%%%%%%%%%%%%%%%%%%%%%%%%%%%%%%%%%%%%%%%%%%%%%%%%%%%%%%
\subsection{Parameter values that maximize the likelihood function in each MM}

\begin{figure}
\centering
\includegraphics[width=\textwidth]{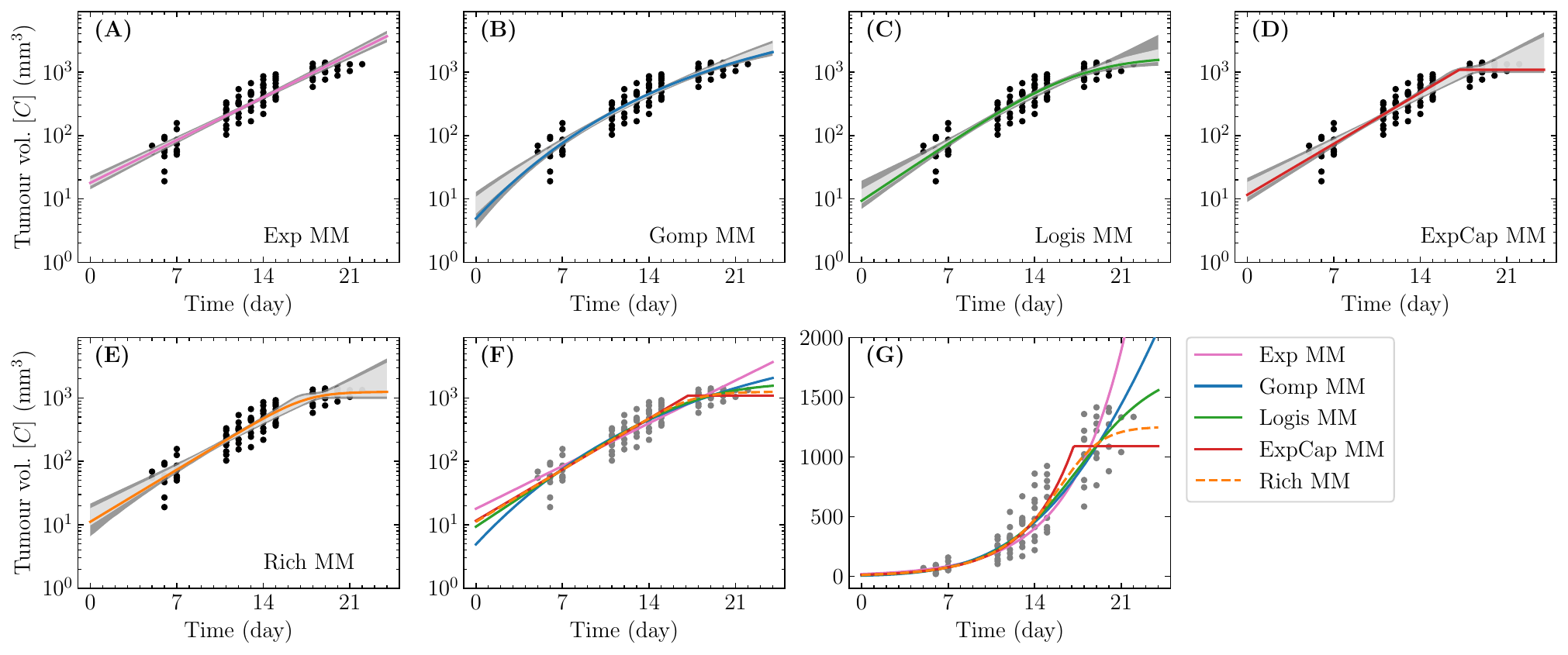} \\
\includegraphics[width=\textwidth]{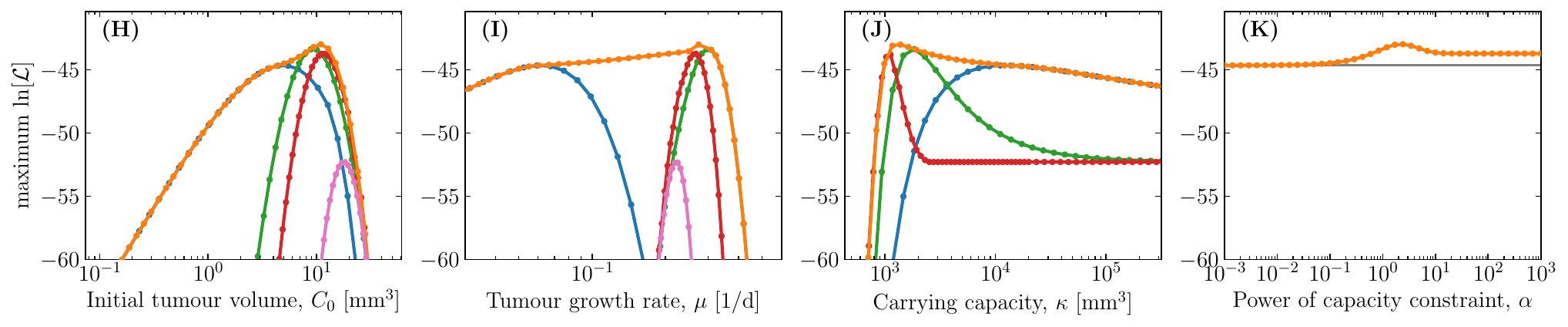}
\caption{%
\textbf{Solutions of each MM fit to the tumour volume measurements and associated parameters.}
(A--E) For each MM explored herein, the curve corresponding to the best-fit parameter set, i.e.\ that which maximizes the likelihood function (Eqn.\ \ref{MMlike}) is showed as a coloured line. A total of 1,000 MCMC-accepted parameter sets were sampled at random with replacement to generate curves of which 68\% (or 95\%) fall within the pale (or dark) grey regions, respectively. (F--G) Tumour volume curve for each MM's best-fit parameters against the data on a logarithmic (F) or linear (G) scale. (H--K) The maximum of the likelihood function (Eqn.\ \ref{MMlike}), as each MM parameter is varied individually.
}
\label{fig:solns_noLODs}
\end{figure}

Figure \ref{fig:solns_noLODs}(A--G) presents the solution of each MM against the data. The variability (68\% and 95\% credible regions) of the MM solutions is the smallest in the intermediate region when all 10 mice have measurable tumour volume, and the growth is largely purely exponential. The solutions varied most, both within each and between MMs, as the number of measurable mice decreased near the lower limit of detection at early times, and at late times when mice were euthanized before the tumour volume exceeded \unit{1,500}{mm^3}.

Figure \ref{fig:solns_noLODs}(H--K) reports the maximum likelihood that is attainable (Eqn.\ \ref{MMlike}) for a given value of each parameter in each of the MMs, sometimes called a profile log-likelihood curve \cite{Raue2009,Eisenberg2017}. For example, given $C_0=\unit{1}{mm^3}$, the value of the remaining parameters is sought so as to achieve the maximum likelihood possible for this value of $C_0$. The tumour in each mouse was initiated with the injection of $10^6$ cells, which is thought to correspond to $C_0\approx\unit{1}{mm^3}$ \cite{benzekry17}. The best-fit (that which maximizes Eqn.\ \ref{MMlike}) initial tumour sizes for all MMs is larger than this estimate, namely from $\sim\unit{5}{mm^3}$ in the \ref{MMgomp}, $\sim\unit{10}{mm^3}$ in the \ref{MMlogis} and \ref{MMgen}, up to $\sim\unit{20}{mm^3}$ in the \ref{MMexp}.

At low $C_0$ values, max.\ $\likeli$ vs $C_0$ curve of the \ref{MMgen} corresponds to that of the \ref{MMgomp}. As $C_0$ is increased, the max.\ \likeli improves and briefly (over a narrow range of $C_0$ values) matches the max.\ \likeli curve of the \ref{MMlogis}, goes on to reach the highest max.\ \likeli, before getting worse as $C_0$ is increased further, eventually approaching the max.\ \likeli curve of the \ref{MMexpcap} and ultimately \ref{MMexp} for the largest $C_0$ values. Note that as different $C_0$ values are explored (i.e., for different assumed values for $C_0$), all other parameters are adjusting accordingly to maximize the \likeli. Since the \ref{MMgen} max.\ \likeli curve for $C_0$ as the latter increases follows the \ref{MMgomp} ($\alpha\to0$), then \ref{MMlogis} ($\alpha=1$), and finally \ref{MMexpcap} ($\alpha\to+\infty$), it suggests that larger values of $C_0$ require larger values of $\alpha$.

In the \ref{MMgen}, a higher \likeli is obtained for $\alpha=1$ \eqref{MMlogis}, than for $\alpha\to0$ \eqref{MMgomp} or $\alpha\to+\infty$ \eqref{MMexpcap}. The best-fit is found for $\alpha\sim2$, but the \likeli varies far less as a function of $\alpha$, than as a function of the other MM parameters. In particular, once $\alpha$ is smaller than $\sim0.1$ (or greater than $\sim10$) making $\alpha$ any smaller (or larger) neither improves nor worsens the \likeli.

\begin{figure}
\centering
\includegraphics[width=\textwidth]{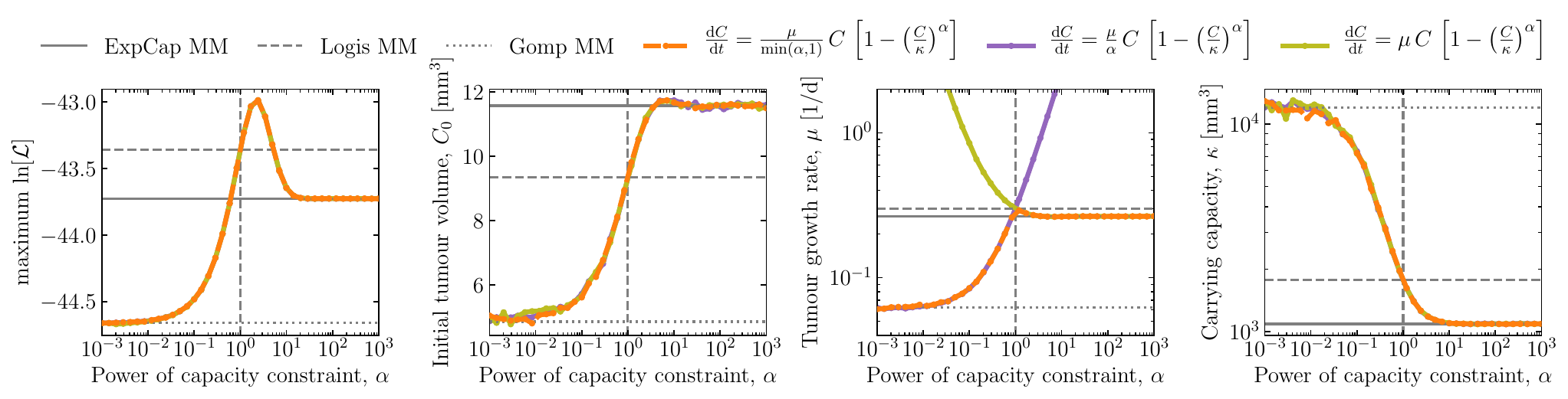}
\caption{%
\textbf{Shift in maximum \likeli parameters as a function of $\alpha$.}
The vertical dashed line indicates $\alpha=1$, corresponding to the \ref{MMlogis}. The horizontal lines correspond to the best-fit value of the quantity on the $y$-axis for the \ref{MMexpcap} ($\alpha\to+\infty$, solid), \ref{MMlogis} ($\alpha=1$, dashed) and \ref{MMgomp} ($\alpha\to 0$, dotted). The 3 coloured curves correspond to variations in the coefficient of the Generalized Logistic ODE: The \ref{MMgen} used herein with a coefficient of $\mu/\min(\alpha,1)$ (orange dashed) is compared to expressions with coefficient $\mu/\alpha$ (purple solid) or simply $\mu$ (olive solid). While coefficient $\mu/\alpha$ yields $\mu\to+\infty$ as $\alpha\to+\infty$ and coefficient $\mu$ yields $\mu\to+\infty$ as $\alpha\to0$, the $\mu/\min(\alpha,1)$ coefficient of the \ref{MMgen} expression used herein provides a more consistent meaning and value for parameters $\mu$ and $\alpha$ as $\alpha$ is varied.
}
\label{fig:compare_alpha}
\end{figure}

Figure \ref{fig:compare_alpha} explores the shift in the best-fit parameters as $\alpha$ is varied in the \ref{MMgen}, and in particular the transition from the \ref{MMgomp} to the \ref{MMlogis} to the \ref{MMexpcap}. The best-fit \ref{MMgen} parameters exactly match those of the \ref{MMgomp} for $\alpha$ less than about 0.1, those of the \ref{MMlogis} at $\alpha=1$, an important transition point for the \ref{MMgen}, and those of the \ref{MMexpcap} for $\alpha$ greater than about 10. As $\alpha$ is varied from small values $\ll1$ to large values $\gg1$, the best-fit values for $C_0$ and $\Cbar$ transition smoothly between the \ref{MMgomp} best-fit values and the \ref{MMexpcap}. There is a relatively narrow range $\sim(0.1,10)$ where the specific value of $\alpha$ has an impact on the max \likeli and the best-fit values of $C_0$ and $\Cbar$.

For parameter $\mu$, the transition in the best-fit value as $\alpha$ is increased is not as simple and it depends on the choice of coefficient for the Generalized Logistic MM. As $\alpha\to0$, $\mu\to+\infty$ if the coefficient of the ODE is $\mu$, or $\mu$ asymptotes to the \ref{MMgomp} best-fit $\mu$ if the coefficient is $\mu/\alpha$, becoming independent of $\alpha$. For $\alpha\to+\infty$, $\mu\to+\infty$ if the coefficient is $\mu/\alpha$, or $\mu$ asymptotes to the \ref{MMexpcap} best-fit $\mu$ if the coefficient is $\mu$. This represents a change in the meaning of parameter $\mu$ and $\alpha$ as $\alpha$ is varied, which is problematic when trying to interpret $\mu$ and $\alpha$ from a biological or physical standpoint. When the coefficient is set to $\mu/\text{min}(1,\alpha)$, as in the \ref{MMgen}, the best-fit $\mu$ gradually increases from the best-fit value in the \ref{MMgomp}, then abruptly settles between the best-fit value for the \ref{MMlogis} and \ref{MMexpcap}, discontinuously rather than smoothly. This provides a more consistent physical meaning for $\mu$ as $\alpha$ transitions from $\alpha<1$ to $\alpha>1$. Replacing $\text{min}(1,\alpha)$ with an expression with a smoother transition around $\alpha=1$, would provide a better, more gradual and consistent meaning for $\mu$ over the full range of the \ref{MMgen} behaviour, but we are unsure how to obtain such an expression.

\begin{figure}
\centering
\includegraphics[width=\textwidth]{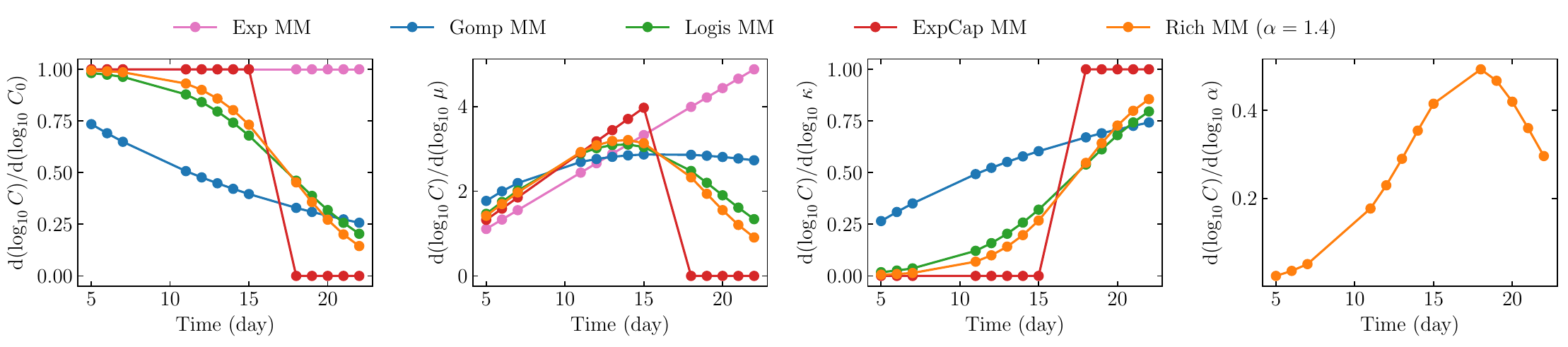}
\caption{%
\textbf{Parameter sensitivity at different measurement time points.}
The central difference of $\log_{10} C$ with respect to the $\log_{10}$ of each parameter about the best-fit parameter set, computed as $\log_{10}[C(t,p\cdot(1+f))/C(t,p\cdot(1-f)]/\log_{10}[(1+f)/(1-f)]$ with $f=10^{-8}$, was used to explore the sensitivity of different measurement time points to relative, percent variations in each parameter.
}
\label{fig:miao_noLOD}
\end{figure}

Figure \ref{fig:miao_noLOD} explores how different measurement time points inform each MM parameter. This parameter sensitivity analysis relies on a local estimation of the derivative of the measurements ($\log_{10} C$) with respect to the $\log_{10}$ of each parameter (relative \% change) about the best-fit parameter set, following a method introduced by Miao et al.\ \cite{miao11}. While $\mu$ is typically best informed by the intermediate time points, in the region where the growth is mostly exponential and all mice have measurable tumour volumes. Unsurprisingly, the initial tumour size, $C_0$, is informed most heavily by the earliest time points, and the carrying capacity, $\Cbar$, by the latest time points.

%%%%%%%%%%%%%%%%%%%%%%%%%%%%%%%%%%%%%%%%%%%%%%%%%%%%%%%%%%%%%%%%%%%%%%%%%%%%%%%%

\subsection{Appropriately handling unmeasurable data points}

In the previous section, all unmeasurable points, i.e.\ those for mice whose tumour was too small to be measured (12 points) and those which had to be euthanized because their tumour had reached a volume of \unit{1,500}{mm^3} (28 points), were neglected. This poses 2 problems. Firstly, this discards meaningful information, namely that the tumour in these mice is known to be either smaller or larger than the lower or upper limit of detection, respectively. Secondly, it introduces a bias in the parameter estimation in that only the largest tumours measurable at early time points, and only the smallest ones at the later time points, are informing the likelihood of parameter sets. This effectively favours a theoretical curve that is higher at early times and smaller at later times than it should otherwise be.

Following the procedure described in \cite{gelmanbayes3} for handling censored data, i.e.\ data with lower and/or upper limits of detection, we revise the likelihood of the data given $\vec{p}$ as
\begin{align}
\mathcal{L}(\mathrm{data}|\pvec) &=
\mathcal{L}_\text{measured}(\mathrm{data}|\pvec) \cdot \mathcal{L}_\text{unmeasured}(\mathrm{data}|\pvec)
\label{MMlikeLOD}
\end{align}
where $\mathcal{L}_\text{measured}(\mathrm{data}|\pvec)$ is that given in Eqn.\ \eqref{MMlike}, and
\begin{align}
\mathcal{L}_\text{unmeasured}(\mathrm{data}|\pvec) &=
\prod_{k=1}^{N_\text{pts}} \left[ 1 +
\frac{1}{2} \mathrm{erf}\left(\frac{\log_{10}[\text{LLD}/C_\text{MM}(\pvec,t_k)]}{\sqrt{2\sigma_C^2}}\right) +
\frac{1}{2} \mathrm{erf}\left(\frac{\log_{10}[C_\text{MM}(\pvec,t_k)/\text{ULD}]}{\sqrt{2\sigma_C^2}}\right) \right]^{U_k} \ .
\label{MMlikeunmeas}
\end{align}
LLD and ULD are the lower and upper limits of detection, respectively, and $U_k$ is the number of mice with unmeasurable tumour volumes at time $t_k$. If all mice had measurable tumour volumes at all time points, i.e.\ if $U_k=0\, \forall\, k$, $\mathcal{L}_\text{unmeasured}(\mathrm{data}|\pvec)=1$ and the likelihood function equals that in Eqn.\ \eqref{MMlike}. But for each time point $t_k$ with a non-zero number of mice with unmeasurable tumour volumes, $U_k> 0$, the expression between the square braces computes the probability (area under the curve) that a tumour volume could be found outside the measurable range, i.e.\ either below the LLD or above the ULD, assuming a normally distributed $\log_{10}$ tumour volume of mean $\log_{10}C_\text{MM}(\pvec,t_k)$ and standard deviation $\sigma_C$. While the ULD is known, namely $\text{ULD} = \unit{1,500}{mm^3}$, the LLD will need to be estimated as an additional parameter, constrained to be $\in\unit{(0,19)}{mm^3}$, where zero is excluded since some mice were unmeasurable and \unit{19}{mm^3} is the smallest tumour volume that was measured within this data set and is therefore measurable. Under these conditions, the most likely value for the LLD is \unit{18.\bar{9}}{mm^3}, i.e.\ the closest value to the maximum allowed value for the LLD.

\begin{figure}
\centering
\includegraphics[width=\textwidth]{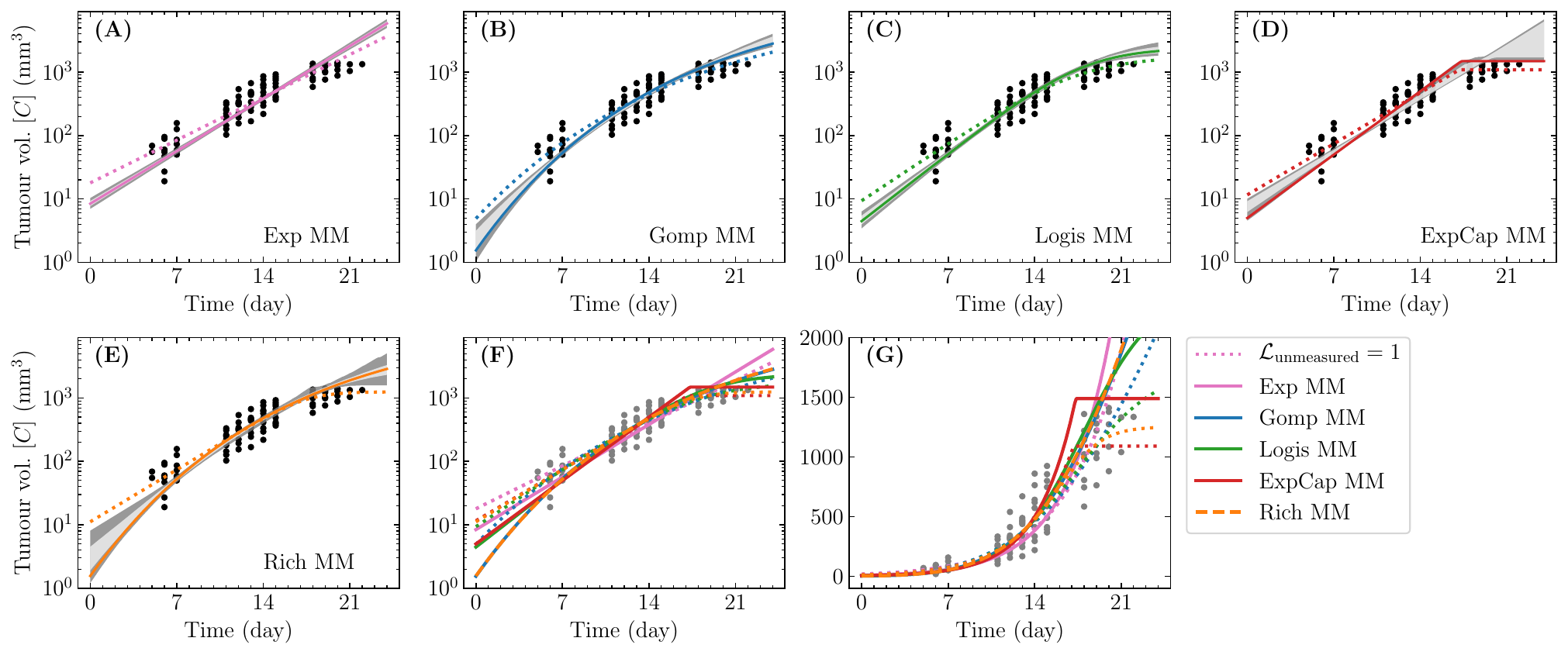}\\
\includegraphics[width=\textwidth]{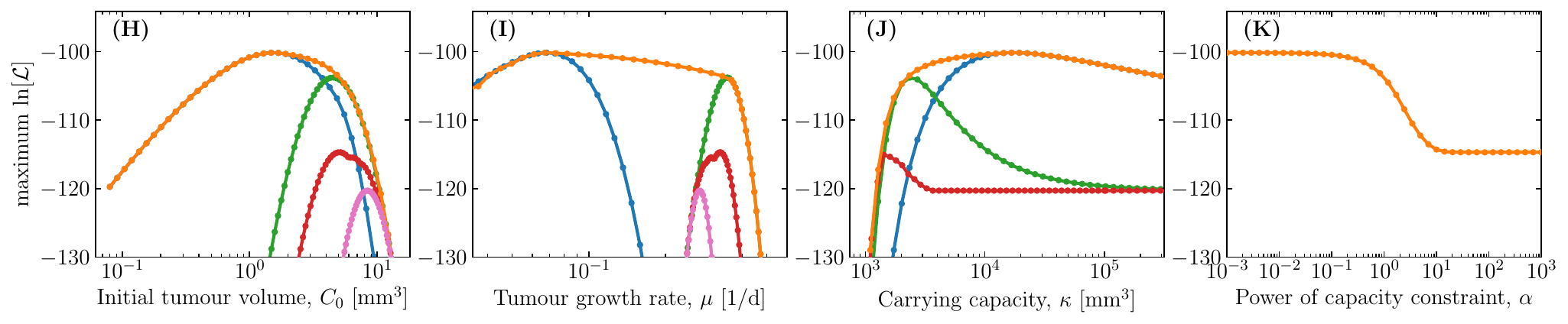}
\caption{%
\textbf{Solutions of each MM and associated parameters when including unmeasurable volumes.}
Panels (A--K) are as in Figure \ref{fig:solns_noLODs}, but correspond to the result obtained when accounting for the 38 unmeasurable tumour volumes by maximizing the likelihood function defined in Eqn.\ \eqref{MMlikeLOD}. For comparison, the best-fit curves when not accounting for unmeasurable data ($\mathcal{L}_\text{unmeasured}=1$, Eqn.\ \eqref{MMlike}) are shown as dashed lines in panels (A--G).
}
\label{fig:solns_LODs}
\end{figure}

Figure \ref{fig:solns_LODs} shows the new solutions and parameters' maximum \likeli curve based on the revised \likeli function which takes into account all 38 time points where the tumour volume was unmeasurable. The new solutions for all 5 MMs are lower at early times and higher at later times, with the old solution (dotted line) falling outside the new solution's 95\% credible region (grey shading), when properly accounting for unmeasurable volumes. Whereas failing to account for the unmeasured tumour volumes favoured a value of $\alpha\sim1.4$, i.e.\ \ref{MMlogis}-like or \ref{MMexpcap}-like rather than \ref{MMgomp}-like, accounting for these unmeasured points favours $\alpha\to 0$, i.e.\ \ref{MMgomp}-like growth.

Consistent with the new solutions predicting lower volumes at earlier times and higher volumes at later times, we find that the best-fit values for the initial tumour volume, $C_0$, are lower and those for the carrying capacity, $\Cbar$, are higher than those of the old solution which did not account for unmeasurable volumes. The best-fit initial tumour volume, $C_0$, is $\sim\unit{1.5}{mm^3}$ for the \ref{MMgomp} and \ref{MMgen}, $\sim\unit{4.5}{mm^3}$ for the \ref{MMlogis} and \ref{MMexpcap}, and $\sim\unit{8.4}{mm^3}$ for the \ref{MMexp}. These values are more consistent with the expected tumour volume given the number of implanted tumour cells, namely $\unit{10^6}{cells} \approx \unit{1}{mm^3}$.

%%%%%%%%%%%%%%%%%%%%%%%%%%%%%%%%%%%%%%%%%%%%%%%%%%%%%%%%%%%%%%%%%%%%%%%%%%%%%%%%

\subsection{Considering the impact of the parameter prior}

Up until now, we have compared MMs and parameter estimation based on maximizing the likelihood function alone. Now we turn our attention to $\prior(\vec{p})$, an important term in Eqn.\ \eqref{MMpost} which, together with the \likeli discussed above, will allow us to estimate the MM parameters' \posterior distribution.

We compare 4 possible \posterior{s}: using the likelihood function given by Eqn.\ \eqref{MMlike} or Eqn.\ \eqref{MMlikeLOD}, combined with either a linearly uniform or a log-uniform prior for all parameters. A linearly uniform prior is simply $\prior(\vec{p}) \propto 1$, namely the $\posterior$ corresponds to the $\likeli$, and a log-uniform prior is given by
\begin{align}
\prior(\vec{p}) \propto \frac{1}{C_0\ \cdot\ \mu\ \cdot\ \Cbar\ \cdot\ \alpha\ \cdot\ \text{LLD}} \ ,
\label{MMprior}
\end{align}
where $\Cbar$ is omitted for the \ref{MMexp}, $\alpha$ is included only for the \ref{MMgen}, and LLD is included only when the \likeli function is given by Eqn. \eqref{MMlikeLOD}, but not when it is given by Eqn.\ \eqref{MMlike}. The MM parameters were constrained to $C_0\in\unit{(0.0,\infty)}{mm^3}$, $\mu\in\unit{(0,\infty)}{1/\dday}$, $\Cbar\in\unit{(0,10^6)}{mm^3}$, $\alpha\in[10^{-5},10^5]$, and $\text{LLD}=\unit{(0,19)}{mm^3}$. The constraint was applied by setting $\prior(\vec{p})=0$ when any parameter in $\vec{p}$ falls outside its bounds.

\begin{figure}
\centering
\includegraphics[width=\textwidth]{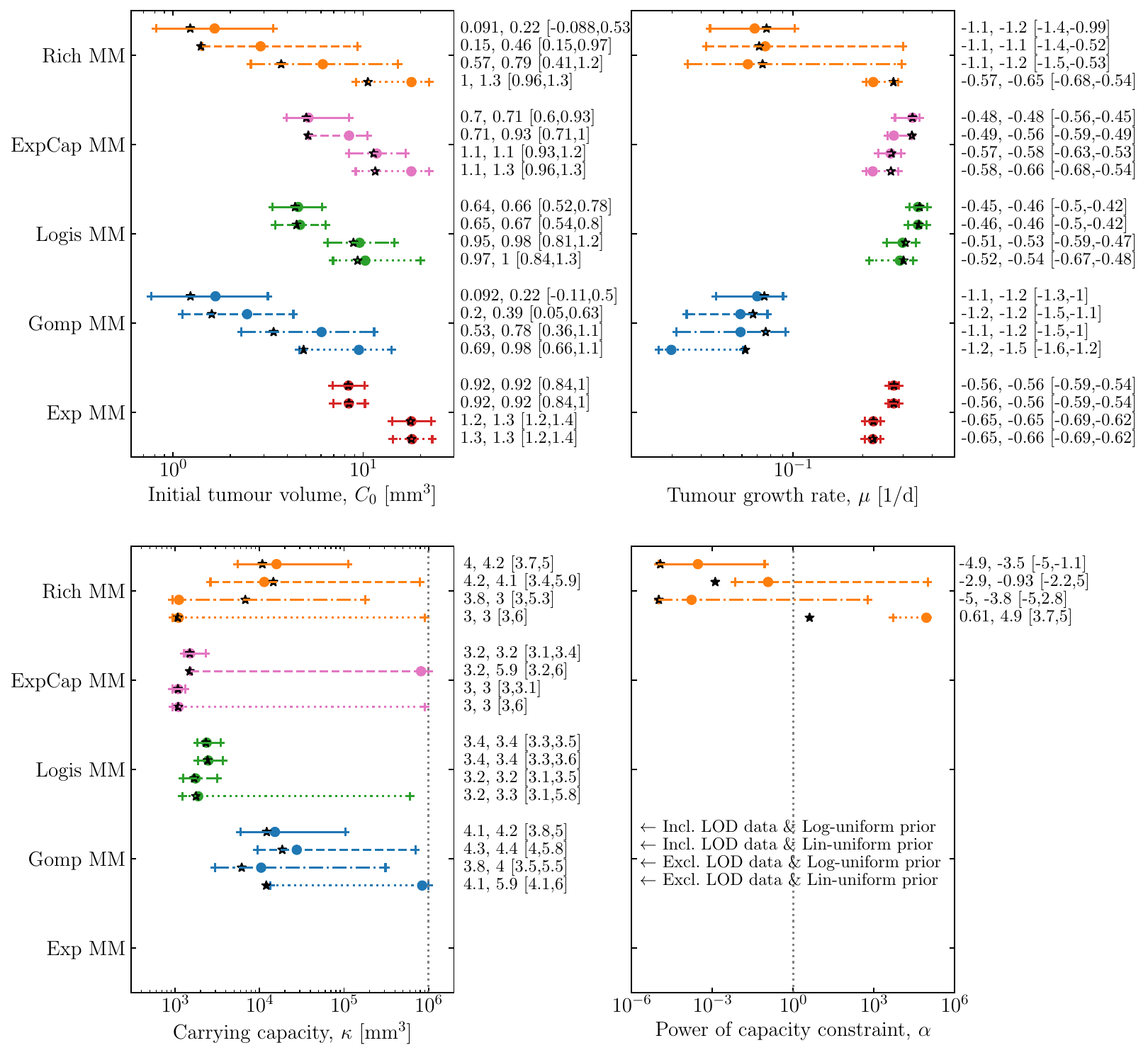}
\caption{%
\textbf{Comparing the parameter posterior distributions under 4 different assumptions.}
For each parameter (different panels), for each MM (different colours) under each of the 4 assumptions (different line styles, annotated in the bottom right panel), the highest-likelihood parameter set ($1^\text{st}$ number, black star) of the multi-dimensional posterior distribution, as well as the mode ($2^\text{nd}$ number, circle) and 95\% CI (numbers in square brace, $+$) for the posterior distribution of each parameter, marginalized over all others, are provided.
}
\label{fig:compare4_params}
\end{figure}

\begin{figure}
\centering
\includegraphics[width=\textwidth]{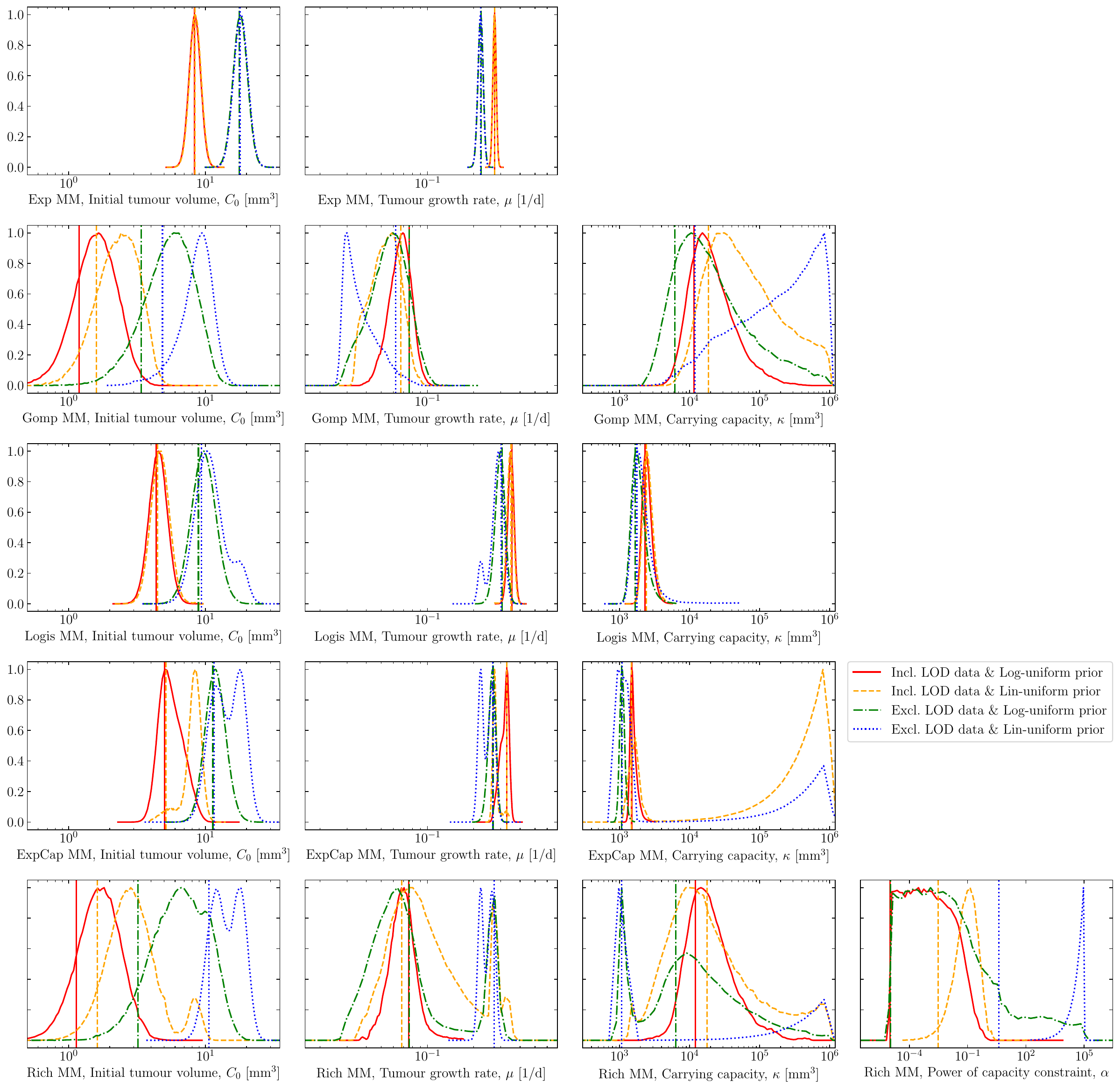}
\caption{%
\textbf{Comparing the parameter posterior distributions under 4 different assumptions.}
For each parameter (different columns), for each MM (different rows) under each of the 4 assumptions (different lines within each graph), the highest-likelihood parameter set (vertical line) of the multi-dimensional posterior distribution, as well as the posterior distribution (curve) of each parameter, marginalized over all others, are provided.
}
\label{fig:compare4_hists}
\end{figure}

Figure \ref{fig:compare4_params} shows the value of each parameter, for each MM, under all 4 scenarios.  Recall that the best-fit parameter value refers to that which maximizes the likelihood function, and the most likely parameter value is that which maximizes the posterior distribution, (i.e.\ the mode). For all MMs, the initial tumour volume ($C_0$) is statistically significantly different under the assumption of a linearly uniform prior when also excluding data outside the LOD (Excl.\ LOD data \& Lin-uniform prior) than under the assumption of a log-uniform prior when including data outside the LOD (Incl.\ LOD data \& Log-uniform prior). In many cases, the inclusion or exclusion of data outside the LOD alone, for a given prior, is sufficient to alter $C_0$ statistically significantly. The impact is less pronounced for the tumour growth rate ($\mu$), and even less so for the carrying capacity ($\Cbar$). In fact, the highest-likelihood and the mode of the $\Cbar$ posterior, marginalized over the other parameters, estimate a larger carrying capacity when including the data outside the LOD than when that data is excluded, for an equivalent prior assumption. Looking at the 95\% CI, however, the shift is not statistically significant in most cases, partly because of the CI's width.

Figure \ref{fig:compare4_hists} provides a more informative view of these results. In particular, it demonstrates why reporting mean or median or highest-likelihood value and 95\% credible interval (CI) can be misleading. Looking at the specific case of the carrying capacity ($\Cbar$), we see that a number of posterior distributions are multi-modal, which translate poorly when reported as a 95\% CI, as in Figure \ref{fig:compare4_params}. Under the assumption of a linearly uniform prior, implying that $\Cbar$ is $20\times$ more likely to be found $\in[10,000, 20,000]$ than $\in[1,000, 2,000]$, we see unsurprisingly that the distribution have a fatter tail at larger values of $\Cbar$, compared to the log-uniform prior, which assumes that $\Cbar$ is equally likely to be found $\in[10,000,20,000]$ or $\in[1,000, 2,000]$. This demonstrates that while this parameter's lower bound is well constrained by the data, this is generally not the case for the upper bound such that the choice of prior significantly impacts the upper portion of the posterior distribution. When including the data beyond the LOD (which one should) and assuming a log-uniform prior (which is more appropriate in our opinion), the posterior distributions are generally better constrained on both sides.

The \ref{MMexp} has the narrowest posterior distributions while the \ref{MMgen} posteriors are the widest, an indication of how well the data can inform MMs with increasing numbers of parameters. As shown in Figure \ref{fig:compare_alpha}, the shift in the maximum likelihood as $\alpha$ is varied from 0 to $+\infty$ is minimal, while the best-fit value of each parameter varies sometimes widely, e.g.\ $\Cbar$ varies from $\sim10^4$ to $10^3$ as $\alpha$ goes from 0 to infinity. This is to be expected since $\alpha$ controls which MM the \ref{MMgen} most resembles, and all the MMs have different posterior distributions for each parameter.

\cleardoublepage
%%%%%%%%%%%%%%%%%%%%%%%%%%%%%%%%%%%%%%%%%%%%%%%%%%%%%%%%%%%%%%%%%%%%%%%%%%%%%%%%
%%%%%%%%%%%%%%%%%%%%%%%%%%%%%%%%%%%%%%%%%%%%%%%%%%%%%%%%%%%%%%%%%%%%%%%%%%%%%%%%
%%%%%%%%%%%%%%%%%%%%%%%%%%%%%%%%%%%%%%%%%%%%%%%%%%%%%%%%%%%%%%%%%%%%%%%%%%%%%%%%
\section{Discussion}

In this work we explored 5 mathematical models (MMs) for cancer tumour growth. We used published data \cite{benzekry14} of Lewis Lung Carcinoma growth in mice to constrain the parameters of the MMs which varied in complexity from 2 to 4 parameters to be estimated. We made use of Bayesian inference to estimate the MM parameters posterior distributions, and made use of the fact that the tumour volume measurement variability between mice appears to follow a log-normal distribution as the basis for our choice of likelihood function.

We demonstrated that including measurements known to be outside the upper and lower limits of detection (LOD), as opposed to simply neglecting them, has an important impact on several parameters. Notably, we showed that by discarding these data, the MM's solutions are biased in such a way that the curves are higher at early times and lower at later times than they are when these data are included. Consequently, the best fit value for the initial tumour volume ($C_0$) shifted to lower values, and that for the carrying capacity or maximum tumour size ($\Cbar$) shifted to higher values, when properly accounting for the censored data. The value of these 2 parameters significantly alters the predicted growth curves beyond the measured time points, which others have relied upon before \cite{vaghi20}.

Many tumour growth data sets are unable to constrain all MM parameters to a satisfying extent, especially in MMs with a larger number of parameters, as was the case here for the Generalized logistic growth MM (\ref{MMgen}), and to a lesser extent the Gompertz and logistic growth MMs. To address this challenge, it is common practice to fix the value of some of the MM parameters. For example, $C_0$ is sometimes fixed based on the number of tumour cells injected into the animal. For the data set used here, this value is said to be $\unit{10^6}{cells}\approx\unit{1}{mm^3}$. By estimating $C_0$ for all our MMs, choosing not to fix any parameter, we found that the best-fit value of $C_0$ was smallest for the Gompertz MM, followed by the logistic and exponential growth MMs. As such, by fixing $C_0$ to a certain value, one favours some MMs at the expense of others, leading to MM selection bias. Since the MM choice also biases the MM-predicted tumour growth curve beyond the measured timepoints, this can be particularly problematic in applications where these MMs are pushed beyond the region of validation.

Generally, we propose an easy to re-use mathematical framework based on Bayes' theorem to estimate MM parameters in a manner that correctly captures inter-individual tumour volume measurement variability and incorporates information from data beyond the LOD. This framework provides not only a robust way to identify the best-fit parameters, i.e.\ those that maximize the likelihood function, but also a parameter and growth curve sensitivity and identifiability analysis in the form of a distribution of solution curves and the shape of the maximum likelihood function about these parameters, sometimes called a profile log-likelihood curve \cite{Raue2009,Eisenberg2017}.

We also showed that the choice of prior when using Bayesian inference has a significant effect on the resulting posterior. This effect is stronger when data is limited \cite{Lambert2005}, as is often the case in mathematical oncology applications. Truncated normal distributions, as in \cite{MacLean2023}, or uniform distributions, as in \cite{Jain2021, TJackson2021,Rocha2022}, are common choices for priors. Guckenberger et al.\ \cite{Guckenberger2013} tested the sensitivity of their posteriors to several choices of prior, including normal, Gamma, and uniform distributions. Log-normal distributions are commonly assumed for model parameters in mathematical oncology, \cite{Jana2022}, to enforce non-negativity in datasets with low mean values and high variances \cite{Keall2006}, such as in tumour measurements \cite{Spratt1969}. However, for some reason, log-normal distributions are not common choices for the prior, despite their use in other fields \cite{Frohlich2019}. In general, a log-normal prior would centre the distribution about measurements of that parameter, but would strongly bias the resulting posterior if data was insufficient to overcome it. Whereas a weakly informative prior, such as log-uniform as used here, reflects a lack of knowledge about the parameter, and relies on the provided data to inform the posterior.

To conclude, we recommend including limit of detection data and log-uniform priors for Bayesian inference methods to estimate model parameters in mathematical oncology applications where data is often limited and noisy. And we hope that this work will encourage others to re-use this framework for other data sets and model parameterization needs.

\cleardoublepage
%%%%%%%%%%%%%%%%%%%%%%%%%%%%%%%%%%%%%%%%%%%%%%%%%%%%%%%%%%%%%%%%%%%%%%%%%%%%%%%%
%%%%%%%%%%%%%%%%%%%%%%%%%%%%%%%%%%%%%%%%%%%%%%%%%%%%%%%%%%%%%%%%%%%%%%%%%%%%%%%%
%%%%%%%%%%%%%%%%%%%%%%%%%%%%%%%%%%%%%%%%%%%%%%%%%%%%%%%%%%%%%%%%%%%%%%%%%%%%%%%%
\section{Acknowledgement}

This work was supported in part by Discovery Grants 355837-2013 and 2022-03744 (CAAB) and 2018-04205 (KPW) from the Natural Sciences and Engineering Research Council of Canada (\url{www.nserc-crsng.gc.ca}), by the Interdisciplinary Theoretical and Mathematical Sciences programme (iTHEMS, \url{ithems.riken.jp}) at RIKEN (CAAB), and by the Toronto Metropolitan University office of the Vice-President, Research and Innovation (JP and DN).

\cleardoublepage
%%%%%%%%%%%%%%%%%%%%%%%%%%%%%%%%%%%%%%%%%%%%%%%%%%%%%%%%%%%%%%%%%%%%%%%%%%%%%%%%
%%%%%%%%%%%%%%%%%%%%%%%%%%%%%%%%%%%%%%%%%%%%%%%%%%%%%%%%%%%%%%%%%%%%%%%%%%%%%%%%
%%%%%%%%%%%%%%%%%%%%%%%%%%%%%%%%%%%%%%%%%%%%%%%%%%%%%%%%%%%%%%%%%%%%%%%%%%%%%%%%

\bibliographystyle{abbrvurl}
\bibliography{sensitivity}

%%%%%%%%%%%%%%%%%%%%%%%%%%%%%%%%%%%%%%%%%%%%%%%%%%%%%%

\end{document}